\documentclass[preprint,aps,pra,groupedaddress,superscriptaddress,floatfix]{revtex4-1}
\usepackage[T2A]{fontenc}
\usepackage[utf8]{inputenc}
\usepackage[english]{babel}
\usepackage{graphicx}
\usepackage{nicefrac}
\usepackage{amsmath}
\usepackage{amsfonts}
\usepackage{amssymb}
\usepackage{amsthm}
\usepackage{epsf}
\usepackage{bm}
\usepackage{longtable}
\newcommand{\phm}{\phantom{-}}

\newcommand{\bfB}{{\bf B}}
\newcommand{\bfF}{{\bf F}}
\newcommand{\bfI}{{\bf I}}
\newcommand{\bfJ}{{\bf J}}

\newcommand{\bfr}{{\bf r}}
\newcommand{\bfp}{{\bf p}}
\newcommand{\balpha}{\bm{\alpha}}

\newcommand{\bra}[1]{\langle#1|}
\newcommand{\ket}[1]{|#1\rangle}

\newcommand{\matrixel}[3]{\langle #1 | #2 | #3 \rangle}
\newcommand{\etal}{\textit{et al.}}
\newcommand{\aZ}{\alpha Z}
\newcommand{\muB}{\mu_\text{B}}
\newcommand{\muN}{\mu_\text{N}}
\newcommand{\masse}{m_\text{e}}
\newcommand{\massp}{m_\text{p}}
\newcommand{\munucl}{\mu^\text{(n)}}
\newcommand{\MJ}{{M_J}}
\newcommand{\Vnucl}{V_\text{nuc}}
\newcommand{\dEmag}{\Delta E_\text{mag}}

\newcommand{\Afns}{A_\text{fns}}
\newcommand{\dghfs}{\delta g_\text{HFS}}
\newcommand{\Dsigmaoph}{\Delta \sigma_\text{1ph}}
\newcommand{\DEoph}{\Delta E_\text{1ph}}
\begin{document}

\title{Nuclear magnetic shielding in highly charged ions}

\author{A.~M.~Volchkova}
\affiliation{Department of Physics, Saint-Petersburg State University, 199034 Saint-Petersburg, Russia}

\author{D.~A.~Glazov}
\affiliation{Department of Physics, Saint-Petersburg State University, 199034 Saint-Petersburg, Russia}

\author{V.~M.~Shabaev}
\affiliation{Department of Physics, Saint-Petersburg State University, 199034 Saint-Petersburg, Russia}

\begin{abstract}
The nuclear magnetic shielding is considered within the fully relativistic approach for the ground state of H-, Li-, and B-like ions in the range $Z=32$--$92$. The interelectronic interaction is evaluated to the first order of the perturbation theory in Li- and B-like ions. The calculations are based on the finite-field method. The numerical solution of the Dirac equation with the magnetic-field and hyperfine interactions included within the dual-kinetic-balance method is employed. The nuclear magnetic shielding constant is an important ingredient for accurate determination of the nuclear magnetic moments from the high-precision $g$-factor measurements.
\end{abstract}
\maketitle
%
%
\section{Introduction}
%
%
Highly charged ions proved to be a versatile tool for tests of the Standard Model, search for new physics, and determination of the fundamental constants and nuclear parameters~\cite{volotka:13:ap,kozlov:18:rmp,shabaev:18:hi,indelicato:19:jpb}. In particular, the nuclear magnetic moments are presently in the focus of interest, since the uncertainty of their determination from the widely accepted nuclear magnetic resonance method may be significantly underestimated~\cite{skripnikov:18:prl,fella:20:prr}. An independent method, free from the environment effects which are difficult to estimate, was proposed in Ref.~\cite{werth:01:ha} and developed further in Ref.~\cite{quint:08:pra}. It is based on the $g$-factor measurement in highly charged ions, where an astonishing progress was achieved during the last two decades~\cite{sturm:17:a}. The relative experimental uncertainty has reached $2.4\times 10^{-11}$ in H-like carbon~\cite{sturm:14:n}, $0.7\times 10^{-10}$ in Li-like silicon~\cite{glazov:19:prl}, and $1.4\times10^{-9}$ in B-like argon~\cite{arapoglou:19:prl}. Determination of the nuclear magnetic moment requires accurate theoretical predictions from the bound-state QED to complement the high-precision measurements. In concert with experiment, relativistic theory of the bound-electron $g$ factor developed remarkably in recent years~\cite{shabaev:15:jpcrd,harman:18:jpcs}. In ions with nonzero nuclear spin, a specific contribution $\dghfs$ arises due to the hyperfine interaction. It can also be expressed in terms of the nuclear magnetic shielding constant $\sigma$. It was investigated in detail within the bound-state QED theory in H-like~\cite{moskovkin:04:pra,yerokhin:11:prl,yerokhin:12:pra} and Li-like~\cite{moskovkin:08:os} ions. The nonlinear effects in magnetic field, described by the Breit-Rabi formula, were considered, in particular, in Refs.~\cite{moskovkin:06:pra,moskovkin:08:pra}.

In the cited works both the hyperfine interaction and the interactions with external magnetic field were considered within the perturbation theory. As an alternative, in Ref.~\cite{volchkova:17:nimb} the so-called finite-field method was used, based on the solution of the Dirac equation, which includes the magnetic-field interaction. The leading contribution to $\sigma$ was evaluated for $1s$, $2s$ and $2p_{1/2}$ states and the full agreement with the perturbative approach was observed. For the ground state of B-like ions, three different effective screening potentials were also considered. Significant screening effect led to conclusion that more rigorous treatment of the interelectronic interaction is in demand. This motivates us to investigate the many-electron effects in these systems. In this work, we present the calculations of the first-order interelectronic-interaction correction, the so-called one-photon exchange, for Li- and B-like ions within the finite-field method. Earlier, this correction was calculated for Li-like ions within the perturbation-theory~\cite{moskovkin:08:os}. The comparison of our results with those from Ref.~\cite{moskovkin:08:os} demonstrates the correctness of the numerical procedure. For B-like ions this term is calculated for the first time.

The relativistic units ($\hbar = 1$, $c = 1$, $\masse = 1$) and the Heaviside charge unit ($\alpha=e^2/(4\pi), e<0$) are used throughout the paper, $Z$ is the nuclear charge, $\muB=|e|/2\masse c$ is the Bohr magneton, $\muN=|e|/2\massp c$ is the nuclear magneton, $\masse$ and $\massp$ are the electron and proton masses, respectively.
%
%
\section{Theory}
%
%
We consider an ion with one electron over the closed shells and nonzero nuclear spin $I$. The energy levels with the electronic angular momentum $J$ are split according to the value of the total angular momentum of the system, $\bfF = \bfJ + \bfI$. The relativistic theory of the hyperfine splitting and the most accurate theoretical values for H-, Li-, and B-like highly-charged ions can be found in Refs.~\cite{shabaev:97:pra,shabaev:98:pra,volotka:08:pra,glazov:19:pra}, see also references therein. In the presence of external magnetic field $\bfB$ (directed along $z$ axis), the ground-state energy level of the ion splits according to the $z$-projection $M_F$ of the total angular momentum $F$. Assuming that the Zeeman splitting is much smaller than the hyperfine splitting, the former can be written as
\begin{equation}
  \dEmag = g_F \muB B M_F
\;,
\end{equation}
where the $g$ factor of the system is given by
\begin{equation}
\label{eq:gF:dg}
  g_F = g_J \frac{\langle \bfJ\cdot\bfF \rangle}{\langle F^2\rangle} - \frac{\masse}{\massp}\,g_I\,\frac{\langle \bfI\cdot\bfF \rangle}{\langle F^2\rangle} + \dghfs
\;,
\end{equation}
with
\begin{align*}
  \langle \bfJ\cdot\bfF \rangle &= \frac{1}{2} [F(F+1)+J(J+1)-I(I+1)]
\;,\nonumber\\
  \langle \bfI\cdot\bfF \rangle &= \frac{1}{2} [F(F+1)+I(I+1)-J(J+1)]
\;,\nonumber\\
  \langle F^2\rangle &= {F(F+1)}
\;.\nonumber
\end{align*}
Here, $g_J$ is the electronic $g$ factor and $g_I$ is the nuclear $g$ factor related to the nuclear magnetic moment via $\munucl = g_I \muN I$. Equation~(\ref{eq:gF:dg}) can be also written as follows
\begin{equation}
\label{eq:gF}
  g_F = g_J \frac{\langle \bfJ\cdot\bfF \rangle}{\langle F^2\rangle} - (1-\sigma)\,\frac{\masse}{\massp}\,g_I\,\frac{\langle \bfI\cdot\bfF \rangle}{\langle F^2\rangle}
\end{equation}
where $\sigma$ is termed as the nuclear magnetic shielding constant. The relation between the hyperfine-interaction correction $\dghfs$ and $\sigma$ reads
\begin{equation}
\label{eq:dghfs}
  \dghfs = \sigma\,\frac{\masse}{\massp}\,g_I\,\frac{\langle \bfI\cdot\bfF \rangle}{\langle F^2\rangle}
\,.
\end{equation}
For the ions under consideration, with one electron above the closed shells, the electronic angular momentum $\bfJ$ is completely determined by the valence electron. In this case, the leading contribution to $\sigma$ (neglecting the interelectronic interaction) can be written as
\begin{equation}
\label{eq:sigma}
  \sigma_0 = \alpha \sum_n \frac{\bra{a} U\ket{n}\bra{n} W\ket{a}}{\epsilon_a-\epsilon_n}
\;,
\end{equation}
where the operators
\begin{equation}
  U = [\bfr \times \balpha]_z
\;,\qquad
  W = \frac{[\bfr\times\balpha]_z}{r^3}
\;,
\end{equation}
originate from the magnetic-field and hyperfine-interaction operators~\cite{volchkova:17:nimb}.
The valence-electron state $\ket{a}$, the intermediate states $\ket{n}$, and their energies $\epsilon_a$ and $\epsilon_n$ are determined from the Dirac equation,
\begin{equation}
\label{eq:D} 
  \left( \balpha \cdot \bfp + \beta + \Vnucl(r) \right) \psi(\bfr) = \epsilon \psi(\bfr) 
\;.
\end{equation}
Here, $\balpha$ and $\beta$ are the Dirac matrices, $\Vnucl(r)$ is the nuclear binding potential. The summation over the spectrum in Eq.~(\ref{eq:sigma}) can be accomplished using the finite-basis-set approach, as was done for H-like~\cite{moskovkin:04:pra} and Li-like~\cite{moskovkin:08:os} ions. For the point nucleus it can also be evaluated analytically~\cite{moore:99:mp,pyper:99:mp1,pyper:99:mp2,moskovkin:04:pra,ivanov:09:pra}.

Alternatively, $\sigma$ can be found by the so-called finite-field method, i.e., from the energies and wave functions, which take into account the magnetic-field and/or hyperfine interactions. In Ref.~\cite{volchkova:17:nimb} only the magnetic-field interaction $\lambda U$ was included in the Dirac equation,
\begin{equation}
\label{eq:D:ffl} 
  \left( \balpha \cdot \bfp + \beta + \Vnucl(r) + \lambda U \right) \psi(\bfr,\lambda) = \epsilon(\lambda) \psi(\bfr,\lambda) 
\;,
\end{equation}
and $\sigma$ was obtained as a derivative of the matrix element $\matrixel{a(\lambda)}{W}{a(\lambda)}$ with respect to $\lambda$,
\begin{equation}
\label{eq:sigma:ffl}
  \sigma_0 = \frac{\alpha}{2}\left.{\frac{\displaystyle {\partial} {\matrixel{a(\lambda)}{W}{a(\lambda)}}}{\displaystyle \partial \lambda}}\right|_{\lambda=0} 
\;.
\end{equation}
In the present work, both magnetic-field and hyperfine interactions are included,
\begin{equation}
\label{eq:D:ff} 
  \left( \balpha \cdot \bfp + \beta + \Vnucl(r) + \lambda U + \mu W \right) \psi(\bfr,\lambda,\mu) = \epsilon(\lambda,\mu) \psi(\bfr,\lambda,\mu) 
\;.
\end{equation}
Then $\sigma_0$ is calculated by taking the mixed derivative,
\begin{equation}
\label{eq:sigma:ff}
  \sigma_0 = \frac{\alpha}{2}\left.{\frac{\displaystyle {\partial}^2 {\epsilon}{(\lambda,\mu)}}{\displaystyle \partial \lambda \displaystyle \partial \mu}}\right|_{\lambda=0,\mu=0} 
\;.
\end{equation}
This approach reduces the summation over the spectrum at the cost of repeating computations for a set of points in $\mu$. For this reason, it is more efficient for higher-order contributions, see the discussion below concerning the one-photon-exchange correction. 

In few-electron systems the interaction between electrons must be taken into account as well. In this work, we consider the first order of the perturbation theory (PT), which is represented by the one-photon-exchange diagrams. This contribution to $\sigma$ was evaluated in Ref.~\cite{moskovkin:08:os} for Li-like ions, where the corresponding diagrams and formulae can be found. In the present work, we consider it within the finite-field approach, i.e., based on the solutions of Eq.~(\ref{eq:D:ff}). The one-photon-exchange correction to the binding energy of the valence electron is given by
\begin{equation}
\label{eq:E-1ph}
  \DEoph = \sum_b \left( \matrixel{a b}{I(0)}{a b} - \matrixel{b a}{I(\epsilon_a-\epsilon_b)}{a b} \right)
\;.
\end{equation}
Here, $\ket{a}$ is the valence-electron state and the summation runs over the closed-shell states $\ket{b}$, the interelectronic interaction operator $I$ in the Feynman gauge is written as
\begin{equation}
  I (\omega,\bfr_{12}) = \alpha (1-{\balpha_1} \cdot \mathbf{\balpha_2}) \,\frac{\exp(i |\omega| r_{12})}{r_{12}}
\;.
\end{equation}
The one-photon-exchange contribution to $\sigma$ can be found from the corresponding energy correction~(\ref{eq:E-1ph}) by taking the mixed derivative,
\begin{equation}
\label{eq:sigma-1ph}
  \Dsigmaoph = \frac{\alpha}{2}\left.{\frac{\displaystyle {\partial}^2\DEoph(\lambda,\mu)}{\displaystyle \partial\lambda \displaystyle \partial\mu}}\right|_{\lambda=0, \mu =0} 
\;.
\end{equation}
Here, it is assumed that $\DEoph$ is calculated by Eq.~(\ref{eq:E-1ph}) with the wave functions $\ket{a}$ and $\ket{b}$ and the energies $\epsilon_a$ and $\epsilon_b$, which correspond to the Dirac equation~(\ref{eq:D:ff}). We note that this expression contains no summations over the spectrum, which normally present in the second and higher orders of the perturbation theory. In particular, the expression for $\Dsigmaoph$ contains double summation within the standard perturbation theory~\cite{moskovkin:08:os}. Within the finite-field approach with single $\lambda$-derivative according to Eqs.~(\ref{eq:D:ffl}) and~(\ref{eq:sigma:ffl}) a single summation would appear.

In order to implement the proposed way to find $\sigma_0$ and $\Dsigmaoph$, we need to solve Eq.~(\ref{eq:D:ff}), which possesses the axial but not the spherical symmetry. The efficient numerical approach to solve the Dirac equation with the axially symmetric potential was developed in Ref.~\cite{rozenbaum:14:pra}. It generalizes the dual-kinetic-balance method~\cite{shabaev:04:prl} with the basis functions constructed from B-splines~\cite{sapirstein:96:jpb}. Initially developed to evaluate the dynamics of an atom in external laser field, it has been applied also to such axially symmetric problems as the Zeeman splitting in highly charged ions~\cite{varentsova:17:nimb,volchkova:17:nimb,volchkova:21}, the electron-positron pair production in heavy-ion collisions~\cite{maltsev:18:pra}, and the binding energies in diatomic molecular ions~\cite{kotov:20:xrs,maltsev:20:os,kotov:21:a}. Below, we briefly recall the basics of this method.

In the axially symmetric potential the $\varphi$-dependence of the Dirac wave function can be written explicitly,
\begin{equation}
\label{eq:psi}
  \psi(\bfr) = \frac{1}{r}
    \begin{pmatrix}
      G_1(r,\theta)\exp{[i(\MJ-\frac{1}{2})\varphi]}\\
      G_2(r,\theta)\exp{[i(\MJ+\frac{1}{2})\varphi]}\\
      iF_1(r,\theta)\exp{[i(\MJ-\frac{1}{2})\varphi]}\\
      iF_2(r,\theta)\exp{[i(\MJ+\frac{1}{2})\varphi]}
    \end{pmatrix}
\;,
\end{equation}
where $\MJ$ is the $z$-projection of the total angular momentum. Equation (\ref{eq:D:ff}) is then reduced to $H_\MJ \Phi = \epsilon \Phi$ with 
\begin{align}
\label{eq:Dr}
  H_\MJ &= 
  \begin{pmatrix}
     1     &  D_\MJ + \tilde D \\
    -( D_\MJ + \tilde D ) & -1 \\
  \end{pmatrix}
\;,\\
  D_\MJ &= \left(\sigma_z \cos{\theta} + \sigma_x \sin{\theta}\right)\left( \frac{\partial}{\partial r} - \frac{1}{r}\right) 
\nonumber\\
    &+ \frac{1}{r}\left(\sigma_x \cos{\theta} - \sigma_z \sin{\theta}\right)\frac{\partial}{\partial \theta} + \frac{1}{r \sin{\theta}}\left( i\MJ \sigma_y + \frac{1}{2} \sigma_x \right)
\;,\\
  \tilde D &= \left( \lambda r + \frac{\mu}{r^2} \right) i \sigma_y \sin\theta
\;,
\end{align}
where the $\tilde D$ term arises from $\lambda U$ and $\mu W$ integrated over $\varphi$ with the exponents from $\psi(\bfr)$~(\ref{eq:psi}), $\sigma_{x,y,z}$ are the Pauli matrices and $\Phi$ is the 4-component wave function, which depends on $r$ and $\theta$,
\begin{equation}
\label{eq:F}
  \Phi(r, \theta) = 
  \begin{pmatrix}
    G_1(r,\theta)\\
    G_2(r,\theta)\\
    F_1(r,\theta)\\
    F_2(r,\theta)
  \end{pmatrix}
\;.
\end{equation}
The functions $G_k(r,\theta)$ and $F_k(r,\theta)$ are decomposed into a finite basis set, constructed from the B-splines in $r$ and the polynomials in $\theta/\pi$. The special dual-kinetic-balance conditions are superimposed on this basis set in order to eliminate the spurious states, see Refs.~\cite{shabaev:04:prl,rozenbaum:14:pra} for details. To find the coefficients of this decomposition and the corresponding energies, the generalized eigenvalue problem is solved numerically. The lowest positive-energy states correspond to the bound states of Eq.~(\ref{eq:D:ff}). The total set of the solutions forms the quasi-complete finite basis set, which effectively represents the infinite spectrum, including the positive- and negative-energy continua, and can be used, in particular, to construct the Green function of Eq.~(\ref{eq:D:ff}).

In order to compute $\sigma_0$ and $\Dsigmaoph$ according to Eqs.~(\ref{eq:sigma:ff}) and (\ref{eq:sigma-1ph}), one has to find the energies $\epsilon_{a,b}(\lambda,\mu)$ and the wave functions $\ket{a}$, $\ket{b}$ at the set of points $\lambda = -n\lambda_0, \dots, -\lambda_0, 0, \lambda_0, \dots, n\lambda_0$ and $\mu = -n\mu_0, \dots, -\mu_0, 0, \mu_0, \dots, n\mu_0$ and use the standard formulae for the derivatives. The proper choice of $\lambda_0$ and $\mu_0$ is very important to obtain the accurate results. A series of calculations for different values of $\lambda_0$, $\mu_0$ and $n$ is performed to choose the optimal values. For the calculations presented in this paper, we use $\lambda_0$ and $\mu_0$ in the range $10^{-4}\dots 10^{-7}$ and $n=1,2,3$.
%
%
\section{Results and discussion}
%
%
In this section, we present the results for the nuclear magnetic shielding constant in the range $Z=32$--$92$, including the one-photon-exchange correction, according to the finite-field method described in detail above. In case there are previously published values, we present them for comparison and observe a good agreement. In order to check additionally the presented method, we calculate also the leading-order hyperfine splitting and the corresponding one-photon-exchange correction. These results are presented in terms of the coefficients $A$ and $B/Z$~\cite{shabaev:97:pra,shabaev:98:pra,volotka:08:pra,glazov:19:pra}. In these papers $A$ is calculated for the point nucleus and the correction due to the finite nuclear size is represented by the factor $1-\delta$. We perform all calculations for the finite nuclear size, hence we denote our value as $\Afns$ and compare it with $A(\aZ)(1-\delta)$ (the Bohr-Weisskopf correction is not taken into account). The values of $B$ from Refs.~\cite{shabaev:98:pra,volotka:08:pra,glazov:19:pra} include the finite-nuclear-size effect. Using the solutions of the Dirac equation~(\ref{eq:D:ff}), these coefficients can be obtained as follows,
\begin{align}
\label{eq:A:ff}
  \Afns &= G_a\left.{\frac{\partial\epsilon(\lambda,\mu)}{\partial\mu}}\right|_{\lambda=0,\mu=0} 
\;,\\
\label{eq:B:ff}
  B/Z &= G_a\left.{\frac{\partial\DEoph(\lambda,\mu)}{\partial\mu}}\right|_{\lambda=0,\mu=0} 
\;,\\
  G_a &= \frac{n^3(2l+1)j(j+1)}{2(\aZ)^3\MJ}
\;,
\end{align}
where $n$, $l$, and $j$ are the quantum numbers of the valence electron. In addition, we calculate $\Afns$ and $B/Z$ by perturbation theory using the finite basis set for Eq.~(\ref{eq:D}) constructed within the dual-kinetic-balance approach~\cite{shabaev:04:prl}.

The nuclear magnetic shielding for the ground $1s$ state of H-like ions was calculated, in particular, in Refs.~\cite{moskovkin:04:pra,yerokhin:11:prl,yerokhin:12:pra} by perturbation theory. In Table~\ref{tab:sigma:H} we compare the results obtained in this work by the finite-field method and those of Ref.~\cite{moskovkin:04:pra}. The values of $\sigma$ are presented in units of $10^{-3}$. The values of the function $S(\aZ)$ from Ref.~\cite{moskovkin:04:pra} are multiplied by the factor $\alpha^2 Z/3$ to obtain $\sigma$. In addition, we compare the results for the leading-order hyperfine-splitting function $\Afns$ with our perturbation-theory calculations and with the corresponding values from Ref.~\cite{shabaev:97:pra} for $Z=82$ and $83$.

In Ref.~\cite{moskovkin:08:os} the $(1s)^2 2s$ ground state of Li-like ions was considered, and the one-photon-exchange correction was calculated. We compare our results for $\sigma_0$ and $\Dsigmaoph$ with Ref.~\cite{moskovkin:08:os} in Table~\ref{tab:sigma:Li}. In this case, the following relations are used: $\sigma_0 = \alpha^2 Z S(\aZ)/12$ and $\Dsigmaoph = \alpha^2 B_{\mu}(\aZ)/12$. The total value of $\sigma = \sigma_0 + \Dsigmaoph$ is also presented in the last column. The results are in good agreement, which can be considered as a demonstration of the correctness of the employed methods. Hyperfine-splitting functions $\Afns$ and $B/Z$ are also presented, the finite-field results are compared with the perturbation theory and with the corresponding values from Ref.~\cite{shabaev:98:pra} for $Z=82$ and $83$.

In Table~\ref{tab:sigma:B} we present our results for the $(1s)^2 (2s)^2 2p_{1/2}$ ground state of B-like ions. The values of $\sigma_0$ are compared with the ones presented in Ref.~\cite{volchkova:17:nimb}. The one-photon-exchange correction is calculated in this work for the first time. We also compare the values of $\Afns$ and $B/Z$ obtained within the finite-field method and by the perturbation theory. For lead and bismuth, the results from Ref.~\cite{glazov:19:pra} are presented as well. We note that the results for B-like ions demonstrate two specific tendencies. First, in contrast to $1s$ and $2s$ states, where $\sigma$ grows fast with $Z$, for $2p_{1/2}$ state $\sigma$ shows roughly $1/Z$-behaviour. A similar feature is observed for the nonlinear Zeeman effect in B-like ions~\cite{lindenfels:13:pra,glazov:13:ps,agababaev:17:nimb,varentsova:17:nimb,varentsova:18:pra}, it is explained by the contribution of the $2p_{3/2}$ state with rather small energy difference in Eq.~(\ref{eq:sigma}). Second, the one-photon-exchange correction is of the same sign as $\sigma_0$ and grows even faster towards low $Z$. So, for low- and middle-$Z$ B-like ions the nuclear magnetic shielding effect is especially strong and the interelectronic-interaction contribution is significant, it should be considered to all orders of the perturbation theory.
%
%
%
\section{Conclusion}
%
%
The nuclear magnetic shielding constant has been evaluated for the ground state of H-, Li-, and B-like ions in the range $Z=32$--$92$ within the fully relativistic approach. For Li- and B-like ions the interelectronic interaction has been taken into account to the first order of the perturbation theory. The finite-field method is used based on the solution of the Dirac equation with the magnetic-field and hyperfine interactions included. The results are compared with the previously published values for H- and Li-like ions. In B-like ions, the one-photon-exchange contribution to the nuclear magnetic shielding has been calculated for the first time. This term is especially important for lower values of $Z$. All-order calculations of the interelectronic interaction for low- and middle-$Z$ ions are in demand. The obtained results can serve for accurate determination of the nuclear magnetic moments from the high-precision $g$-factor measurements anticipated in the near future.
%
%
\acknowledgments
%
%
The work was supported by RFBR (Grants No.~19-32-90278 and No.~19-02-00974).
%
%
%

%
\newpage
%
%
%
%
\setlength{\tabcolsep}{12pt}

\begin{table}

\caption{ 
\label{tab:sigma:H} 
The hyperfine-splitting factor $\Afns$ and the nuclear magnetic shielding constant $\sigma_0$ for the ground $1s$ state of H-like ions. For $\Afns$ the results of our perturbation-theory calculations and of Ref.~\cite{shabaev:97:pra} are given for comparison. The values of $\sigma_0=\alpha^2 Z S(\aZ)/3$ from Ref.~\cite{moskovkin:04:pra} are also presented. 
The values of $\sigma$ are given in units of $10^{-3}$}
\begin{center}
\begin{tabular}{rll}
\hline
 $Z$
& $\Afns$
& $\sigma_0 \times 10^3$
\\
\hline
$32$ & $1.0813$     & $0.6578$   \\
     & $1.0813^a$   & $0.6579^b$ \\
\hline
$54$ & $1.2687$     & $1.4615$   \\
     & $1.2686^a$   & $1.4617^b$ \\
\hline
$82$ & $1.8537$     & $3.9548$   \\
     & $1.8548^a$   & $3.9579^b$ \\
     & $1.8545^c$   &            \\
\hline
$83$ & $1.8881$     & $4.1093$   \\
     & $1.8892^a$   & $4.1109^b$ \\
     & $1.8889^c$   &            \\
\hline
$90$ & $2.1688$     & $5.397$    \\
     & $2.1694^a$   &            \\
\hline
$92$ & $2.2635$     & $5.843$   \\
     & $2.2653^a$   & $5.851^b$ \\
\hline
\end{tabular}
\end{center}
$^a$perturbation theory, \qquad
$^b$Moskovkin~\etal~\cite{moskovkin:04:pra}, \qquad
$^c$Shabaev~\etal~\cite{shabaev:97:pra}
\end{table}
\begin{table}
\caption{ 
\label{tab:sigma:Li} 
The hyperfine-splitting coefficients ($\Afns$ and $B/Z$) and the nuclear magnetic shielding constant ($\sigma_0$, $\Dsigmaoph$, and $\sigma = \sigma_0 + \Dsigmaoph$) for the ground $(1s)^2 2s$ state of Li-like ions. For $\Afns$ and $B/Z$ the results of our perturbation-theory calculations and of Ref.~\cite{shabaev:98:pra} are given for comparison. The results for $\sigma_0=\alpha^2 Z S(\aZ)/12$ and $\Dsigmaoph=\alpha^2 B_{\mu}(\aZ)/12$ from Ref.~\cite{moskovkin:08:os} are also presented. The values of $\sigma$ are given in units of $10^{-3}$}
\begin{center}
\begin{tabular}{rlllll}
\hline
 $Z$
& $\Afns$
& $B/Z$
& $\sigma_0 \times 10^3$
& $\Dsigmaoph \times 10^3$
& $\sigma \times 10^3$
\\
\hline
$32$ & $1.1196$    & $-0.0940$    & $0.155542$   & $-0.008312$   & $0.147230$ \\
     & $1.1196^a$  & $-0.0941^a$  & $0.155586^b$ & $-0.008325^b$ & $0.147261$ \\
\hline
$54$ & $1.4062$    & $-0.0718$    & $0.318991$   & $-0.011416$   & $0.307575$ \\
     & $1.4061^a$  & $-0.0718^a$  & $0.319111^b$ & $-0.011416^b$ & $0.307695$ \\
\hline
$82$ & $2.3958$    & $-0.0858$    & $0.80409$    & $-0.02303$    & $0.78106$  \\
     & $2.3973^a$  & $-0.0859^a$  & $0.80556^b$  & $-0.02303^b$  & $0.78254$  \\
     & $2.3987^c$  & $-0.0856^c$  &              &               &            \\
\hline
$83$ & $2.4579$    & $-0.0872$    & $0.83577$    & $-0.02377$    & $0.8120$ \\
     & $2.4595^a$  & $-0.0873^a$  & $0.83650^b$  & $-0.02379^b$  & $0.8127$ \\
     & $2.4609^c$  & $-0.0870^c$  &              &               &            \\
\hline
$90$ & $2.9795$    & $-0.0998$    & $1.1013$     & $-0.03031$    & $1.0710$ \\
     & $2.9803^a$  & $-0.0998^a$  & $1.1040^b$   & $-0.03036^b$  & $1.0736$ \\
\hline
$92$ & $3.1613$    & $-0.1044$    & $1.1960$     & $-0.03262$    & $1.1633$ \\
     & $3.1641^a$  & $-0.1044^a$  & $1.1983^b$   & $-0.03263^b$  & $1.1657$ \\
\hline
\end{tabular}
\end{center}
$^a$perturbation theory, \qquad
$^b$Moskovkin~\etal~\cite{moskovkin:08:os}, \qquad
$^c$Shabaev~\etal~\cite{shabaev:98:pra}
\end{table}
\begin{table}
\caption{ 
\label{tab:sigma:B} 
The hyperfine-splitting coefficients ($\Afns$ and $B/Z$) and the nuclear magnetic shielding constant ($\sigma_0$, $\Dsigmaoph$, and $\sigma = \sigma_0 + \Dsigmaoph$) for the $(1s)^2 (2s)^2 2p_{1/2}$ ground state of B-like ions. For $\Afns$ and $B/Z$ the results of our perturbation-theory calculations and of Ref.~\cite{glazov:19:pra} are given for comparison. The results for $\sigma_0$ from Ref.~\cite{volchkova:17:nimb} are also presented. The values of $\sigma$ are given in units of $10^{-3}$}
\begin{center}
\begin{tabular}{rlllll}
\hline
 $Z$
& $\Afns$
& $B/Z$
& $\sigma_0 \times 10^3$
& $\Dsigmaoph \times 10^3$
& $\sigma \times 10^3$
\\
\hline
$32$ & $1.11652$    & $-0.20966$   & $4.7447$     & $\phm0.3522$  & $5.0969$    \\
     & $1.11652^a$  & $-0.20949^a$ & $4.745^b$    &               &             \\
\hline
$54$ & $1.39693$    & $-0.17115$   & $2.9612$     & $\phm0.1152$  & $3.0764$    \\
     & $1.39693^a$  & $-0.17096^a$ &              &               &             \\
\hline
$82$ & $2.43299$    & $-0.24117$   & $2.2610$     & $\phm0.0226$  & $2.2836$    \\
     & $2.43338^a$  & $-0.24085^a$ & $2.260^b$    &               &             \\
     & $2.43344^c$  & $-0.23799^c$ &              &               &             \\
\hline
$83$ & $2.50186$    & $-0.24729$   & $2.2513$     & $\phm0.0243$  & $2.2756$    \\
     & $2.50229^a$  & $-0.24697^a$ & $2.252^b$    &               &             \\
     & $2.50229^c$  & $-0.24614^c$ &              &               &             \\
\hline
$90$ & $3.1050$     & $-0.3032$    & $2.2319$     & $\phm0.0018$  & $2.2337$   \\
     & $3.1052^a$   & $-0.3028^a$  &              &               &            \\
\hline
$92$ & $3.3262$     & $-0.3246$    & $2.2403$     & $-0.0024$     & $2.2379$   \\
     & $3.3271^a$   & $-0.3241^a$  & $2.240^b$    &               &            \\
\hline
\end{tabular}
\end{center}
$^a$perturbation theory, \qquad
$^b$Volchkova~\etal~\cite{volchkova:17:nimb}, \qquad
$^c$Glazov~\etal~\cite{glazov:19:pra}
\end{table}
\end{document}